# Confirmation of Lagrange Hypothesis for Twisted Elastic Rod

V. Kobelev

**Abstract**.
The history of structural optimization as an exact science begins possibly with the celebrated Lagrange problem: to find a curve which by its revolution about an axis in its plane determines the rod of greatest efficiency [1]. The Lagrange hypothesis, that the optimal rod possesses the constant cross-section was abandoned for Euler buckling problem [2]. In this Article the Lagrange hypothesis is proved to be valid for Greenhill's problem of torque buckling. The corresponding isoperimetric inequality is affirmed.

## 1. Introduction

We study the optimal shape of a thin elastic rod, twisted by couples applied at its ends alone (Greenhill's problem). The ends of the rod are assumed to be attached to the supports by ideal spherical hinges and are free to rotate in any directions. The twisting couple retains its initial direction during buckling. The distribution of material along the length of a twisted rod is optimized so that the rod is of minimum volume and will support a given moment without spatial buckling. Introduction of new unknown variables reduces the problem to the exactly solvable auxiliary problem with constant coefficients, such that the critical buckling moment of a rod with an arbitrary shape allows the exact representation as the certain integral functional. The strict upper boundary of this functional is stated by means of Hölder's inequality. Thus, the exact solution of optimization problem is stated in form of isoperimetric inequality, involving the length of the rod, its volume and critical torque. Remarkable, that in accordance with Lagrange hypothesis in the torsion problem the optimal shape of the rod is constant along its length.

## 2. Exact solution of the Greenhill's problem with an isotropic cross-section

Consider a thin elastic rod with isotropic cross-section, twisted by couples applied at its ends alone (Greenhill's problem). The ends of the rod $x=0$ and $x=l$ are assumed to be attached to the supports by ideal spherical hinges and are free to rotate in any directions. The twisting couple $M$ retains its initial direction during buckling. The equations of rod with constant bending stiffness $EJ$ over the length read

(1) $EJ\dfrac{d^2 y}{dx^2} = M\dfrac{dz}{dx}, \qquad EJ\dfrac{d^2 z}{dx^2} = -M\dfrac{dy}{dx},$

(2) $y(0) = z(0) = y(l) = z(l) = 0$.

The integration of the equations (1) delivers new equations with two unknown constants $c_1, c_2$:

(3) $EJ \dfrac{dy}{dx} = Mz + c_1, \qquad EJ \dfrac{dz}{dx} = -My + c_2.$

The problem (1)-(2) and its equivalent problem (2),(3) are neither selfadjoint nor conservative in the classical sense. However, it is a conservative system of the second kind [3]. From earlier investigation [4] it is known that indeed the trace of the eigenvalue curve is such that the system buckles by divergence, and the that the exact buckling loads of Greenhill's problem is

(4) $M^* = \dfrac{2\pi EJ}{l}.$

Introduction of the independent variable $\xi$ [5]

(5) $x = \int\limits_0^\xi \dfrac{dt}{F(\xi)}, \qquad F(\xi) > 0,$

reduces the problem (2)-(3) to

(6) $F(\xi)EJ \dfrac{dY}{d\xi} = MZ + c_1,\ F(\xi)EJ \dfrac{dZ}{d\xi} = -MY + c_2,\ Y(0) = Z(0) = Y(L) = Z(L) = 0$

with

$y(x) = y\left(\int\limits_0^\xi \dfrac{dt}{F(t)}\right) = Y(\xi),\ z(x) = z\left(\int\limits_0^\xi \dfrac{dt}{F(t)}\right) = Z(\xi)$

The value $L$ is the solution of the algebraic equation

(7) $l = \int\limits_0^L \dfrac{d\xi}{F(\xi)}.$

Eigenvalue problem (5)-(6) describes the Greenhill's problem with a variable bending stiffness $EJF(\xi)$ over the length. The critical eigenvalues of the problems (5)-(6) and (2)-(3) match and are equal to the value (4). Substitution of (7) in the expression (4) delivers the analytical expression for critical torque

(8) $M^* = M^*[F(t)] = \dfrac{2\pi E}{\int\limits_0^L \dfrac{dt}{F(t)J}}.$

This integral functional delivers the critical load of Greenhill's problem with a variable cross-section of the length $L$ in terms of shape function $F(t) > 0$.

### 3. Optimization problem and isoperimetric inequality

The volume of the rod is given

(9) $V = \int\limits_0^L A(t)dt.$

We assume that second moment of the $F(\xi)J$ and the cross-sectional area $A(\xi)$ are related by

(10) $\quad F(\xi)J = \alpha_n A^n(\xi),$

where $n = 1,2,3$ and $\alpha_n$ is a constant that depends on $n$. For example, for $n = 2$ the $\alpha_2 = (4\pi)^{-1}$.

We define the optimal twisted rod as the rod so shaped that any other rod of the same length and the same volume will buckle under the smaller critical load. The formal formulation of the optimization problem is the following:

$$M^{**} = \max_{F(\xi)} M^*[F(\xi)], \qquad V = const.$$

We demonstrate now, that the optimal twisted rod possess the constant cross-section over the span length, such that $F(\xi) = 1$.

For this purpose we use the Hölder inequality [6]. If $f(\xi) \geq 0, g(\xi) \geq 0, 0 \leq \xi \leq L$ are measurable real- or complex-valued functions, then Hölder inequality is

(11) $\quad \int_0^L f(\xi)g(\xi)d\xi \leq \left(\int_0^L f^p(\xi)d\xi\right)^{1/p} \left(\int_0^L g^q(\xi)d\xi\right)^{1/q}.$

The numbers $p$ and $q$ above are said to be Hölder conjugates of each other

(12) $\quad \dfrac{1}{p} + \dfrac{1}{q} = 1. \qquad 1 \leq p, q \leq \infty.$

Substitution of the following expressions

(13) $\quad f(\xi) = A^{\frac{1}{1+n}}(\xi), \qquad g(\xi) = A^{\frac{-1}{1+n}}(\xi), \qquad \dfrac{1}{p} = \dfrac{1}{n+1}, \dfrac{1}{q} = \dfrac{n}{n+1}$

into Hölder inequality (11) leads immediately to the desired isoperimetric inequality. Namely,

$$\int_0^L f^p(\xi)d\xi = \int_0^L A(\xi)d\xi \equiv V, \quad \int_0^L g^q(\xi)d\xi = \int_0^L \dfrac{d\xi}{A^n(\xi)} \equiv \alpha_n \int_0^L \dfrac{d\xi}{JF(\xi)}, \quad \int_0^L f(\xi)g(\xi)d\xi = L.$$

With these integrals the Hölder inequality (11) reads

(14) $\quad L \leq V^{\frac{1}{n+1}} \left(\alpha_n \int_0^L \dfrac{d\xi}{F(\xi)J}\right)^{\frac{n}{n+1}}.$

Substitution (8) into (14) states the upper boundary $M^{**}$ for critical buckling moment in form of isoperimetric inequality

(15) $\quad M^* \leq M^{**} \equiv 2\pi E \alpha_n \left(\dfrac{V}{L^{n+1}}\right)^{1/n}.$

The equality attains only for the rod with constant cross-section over the span length. With the inequality (14) we prove for Greenhill's buckling problem the Lagrange hypothesis, that the rod with constant cross-section is optimal.

## 4. Twist buckling of rod with an anisotropic cross-section

The solution of buckling problem could be extended for a thin elastic strut with anisotropic cross-section. We assume that the moments of inertia of cross-section $J_y$, $J_z$ with respect to axis $y$ and $z$ are different, but the ratio $k = J_y / J_z$ keeps constant along the length of the rod.

The eigenvalue problem reads

(16) $\quad F(\xi)EJ_z \dfrac{dY}{d\xi} = MZ + c_1, \ F(\xi)EJ_y \dfrac{dZ}{d\xi} = -MY + c_2,$

$Y(0) = Z(0) = Y(L) = Z(L) = 0$.

The substitution

$Y(\xi) = k^{1/4}\, \tilde{Y}(\xi), \ Z(\xi) = k^{-1/4}\, \tilde{Z}(\xi)$

reduces the problem (16) to its isotropic formulation in terms of $\tilde{Y}(\xi)$, $\tilde{Z}(\xi)$ with $J = \sqrt{J_y J_z}$. The corresponding isoperimetric inequality remains its validity.